\documentclass[journal=jacsat,manuscript=article,superscriptaddress]{achemso}

\usepackage[version=3]{mhchem} 
\usepackage{booktabs}
\usepackage{colortbl}
\usepackage[table]{xcolor}
\usepackage{siunitx}

\author{Gabriele Amante}
\affiliation{Institute for Chemical-Physical Processes, National Research Council (CNR-IPCF), V.le F. Stagno d'Alcontres 37, 98158 Messina, Italy}
\alsoaffiliation{Department of Chemical, Biological, Pharmaceutical, and Environmental Sciences, University of Messina, Viale F. Stagno d’Alcontres 31, 98166 Messina, Italy}

\author{Klaudia Mráziková}
\affiliation{Institute for Chemical-Physical Processes, National Research Council (CNR-IPCF), V.le F. Stagno d'Alcontres 37, 98158 Messina, Italy}
\alsoaffiliation{J. Heyrovský Institute of Physical Chemistry, Academy of Sciences of the Czech Republic, Dolejškova 3, CZ-18223 Prague 8, Czech Republic}

\author{Gabriele Centi}
\affiliation{Department of Chemical, Biological, Pharmaceutical, and Environmental Sciences, University of Messina, Viale F. Stagno d’Alcontres 31, 98166 Messina, Italy}

\author{Sylvie Roke}
\affiliation{Laboratory for Fundamental BioPhotonics (LBP), Institute of Bioengineering, École Polytechnique Fédérale de Lausanne (EPFL), Lausanne, Switzerland}

\author{Ali Hassanali}
\email{ahassana@ictp.it}
\affiliation{International Centre for Theoretical Physics (ICTP), Strada Costiera 11, 34151 Trieste, Italy}

\author{Giuseppe Cassone}
\email{giuseppe.cassone@cnr.it}
\affiliation{Institute for Chemical-Physical Processes, National Research Council (CNR-IPCF), V.le F. Stagno d'Alcontres 37, 98158 Messina, Italy}

\title[Charge transfer at oil-water boundaries]
{Collective Electronic Polarization Drives Charge Asymmetry at Oil-Water Interfaces}

\abbreviations{IR,NMR,UV}

\begin{document}

\begin{tocentry}

Some journals require a graphical entry for the Table of Contents.
This should be laid out ``print ready'' so that the sizing of the
text is correct.

Inside the \texttt{tocentry} environment, the font used is Helvetica
8\,pt, as required by \emph{Journal of the American Chemical
Society}.

The surrounding frame is 9\,cm by 3.5\,cm, which is the maximum
permitted for  \emph{Journal of the American Chemical Society}
graphical table of content entries. The box will not resize if the
content is too big: instead it will overflow the edge of the box.

This box and the associated title will always be printed on a
separate page at the end of the document.

\end{tocentry}

\begin{abstract}
 Why kinetically stable oil droplets in water spontaneously acquire a negative charge remains one of the most vigorously debated questions in interfacial science. Here, we combine neural-network based deep potential molecular dynamics with a data-driven and information theory approach to probe the real-space electron density at an extended decane-water interface. While decane-water clusters show nearly symmetric forward and backward charge transfer (CT) and thus negligible net CT, the extended interface displays a systematic electronic asymmetry, yielding a net CT from water to the hydrocarbon phase producing an average surface charge density of $\sim0.006~e^{-}\,\mathrm{nm}^{-2}$ on the oil phase. This imbalance is accompanied by much larger intra-phase self-polarization, particularly within the hydrocarbon phase, demonstrating that collective many-body polarization dominates the interfacial electronic response. Structural analysis reveals an asymmetry between forward C--H$\cdots$O and backward O--H$\cdots$C motifs, providing a microscopic origin for a net CT from one phase to the other. Curiously, both the water O--H and decane C--H covalent bonds incur subtle contractions which originate from a response to the charge-separation layers at the interface. These features are fully consistent with the weak improper hydrogen-bonds forming at the oil-water interface that results in blue-shifts of the C-H modes.
\end{abstract}


\section{Introduction}

Oil-water interfaces play a central role in a wide range of phenomena spanning soft-matter physics\cite{Kim2022,clegg2024}, colloid science\cite{dietrich2008,zuo2022}, atmospheric chemistry\cite{faye2017,limmer2024}, and microdroplet reactivity\cite{vogel2020,burkhard2024,ashok2025,zheng2025}. A particularly intriguing and long-standing observation is that oil droplets dispersed in water acquire a sizable negative $\zeta$-potential even in the absence of added surfactants or electrolytes. This charge stabilizes emulsions against coalescence\cite{Pullanchery2021Science} and drives the electrophoretic mobility of oil droplets in the presence of external electric fields\cite{agmon2016}. Recent spectroscopic measurements inferred the existence of exceedingly large interfacial electric fields ($\sim$50-90 MV cm$^{-1}$) associated with the droplet $\zeta$-potential which have been proposed to be important for enhanced reactivity at interfaces.\cite{Shi2025Nature} Because the $\zeta$-potential exhibits a clear pH dependence, the negative charge has traditionally been attributed to hydroxide ions (OH$^-$) adsorbing at the hydrophobic interface. 

Direct experimental and theoretical evidence for hydroxide accumulation at such interfaces has remained elusive below bulk concentrations of $\sim1$ M (pH 14). Several label-free, surface-specific spectroscopic methods designed to probe hydroxide directly -- such as charge-transfer-to-solvent (CTTS) transitions\cite{PetersenCPL2008}, 2p-$\pi$ electronic transitions\cite{HermanssonCPL2011}, and the O-H$^-$ vibrational stretch\cite{TianJACS2008,TianPNAS2009,YangPRL2020,StrazdaiteJCP2015,Vacha2011JACS,PullancheryNatCommun2024} -- do not reveal any pH-dependent interfacial accumulation of OH$^-$ below pH $\sim13-14$ (corresponding to $\sim100$ mM--1 M). Similarly, most molecular simulations suggest that hydroxide ions are repelled from hydrophobic surfaces or at best, too weakly attracted to account for the large negative zeta potentials\cite{agmon2016,baermundy2014,voth2015,jansjavkatnetz2017,Zhang_Langmuir25}. This has thus motivated the search for alternative explanations for the origin of the interfacial charge.

One frequently invoked mechanism attributes the negative charge of oil droplets to charge transfer (CT) at the oil-water interface. Evidence for this mechanism has primarily been motivated from electronic-structure calculations of the water dimer\cite{khaliulin2009} combined with molecular dynamics simulations of extended interfaces, which reveal pronounced hydrogen-bond asymmetries that are typically enhanced at water surfaces\cite{Vacha2011JACS,jungwirth2012,Poli2020NatCommun}. While early studies focused on CT between water molecules, Poli and co-workers proposed that oriented interfacial water molecules could donate electronic charge to hydrocarbon groups\cite{Poli2020NatCommun}, thereby generating a negatively charged oil phase. Subsequent surface-sensitive vibrational spectroscopy measurements revealed distinct interfacial signatures in both the O-H and C-H stretching regions. These features were interpreted as evidence for \emph{improper} C--H···O hydrogen bonds that could facilitate CT from water to the hydrocarbon phase.\cite{Pullanchery2021Science} 



The magnitude and spatial extent of CT, however, depend sensitively on both the level of electronic-structure theory and the scheme used to partition the electron density\cite{Poli2020NatCommun}. In this context, recent energy-decomposition analyses of water-hydrocarbon clusters have argued that forward (water$\rightarrow$oil) and backward (oil$\rightarrow$water) CT contributions largely cancel, resulting in negligible net CT and attributing spectroscopic shifts primarily to Pauli repulsion and polarization effects\cite{Zhao2025Angew}. While such cluster-based approaches provide valuable insight into individual intermolecular interactions, they are intrinsically limited in their ability to capture the collective, many-body, and non-local polarization effects that emerge only at extended interfaces.

Here, we revisit the molecular and electronic origins of CT at extended oil-water interfaces by combining Deep Potential Molecular Dynamics (DPMD) simulations of a decane-water interface with first-principles electronic-structure calculations. By explicitly probing how the real-space electron density responds to interface formation using an information-theoretic analysis, we obtain an atom-resolved picture of interaction-induced charge rearrangements. This framework allows us to disentangle inter-phase CT from intra-phase polarization and directly quantify both forward and backward CT processes at the interface. Our results reveal that a net transfer of electronic charge from water to the oil phase emerges only at the extended interface and is not captured by isolated cluster models. The resulting charge redistribution is accompanied by structural distortions of the interfacial hydrocarbon groups that are consistent with the experimentally observed blue shift of interfacial C-H vibrational modes, providing an unprecedented direct molecular-level picture of interaction-induced charge rearrangements at oil-water interfaces.

\section{Methods}

\subsection{Deep Potential Molecular Dynamics and Electronic Structure Calculations}

With the aim of exploring the configurational space accessible to an extended oil-water interface, a Deep Potential Molecular Dynamics (DPMD) simulation of a decane-water slab was carried out with the LAMMPS software (v. 29Aug2024) \cite{LAMMPS}. The reactive DP trained on the hybrid meta-GGA M06-2X \cite{M062X} exchange and correlation functional reported in Ref. \cite{Zhang_Langmuir25} was employed. The simulation box was composed of 15 decane (C$_{10}$H$_{22}$) and 156 water (H$_2$O) molecules (\emph{i.e.}, 948 atoms) arranged in a parallelepiped box of $a=b=19.641$ {\AA} and $c=25.000$ {\AA} yielding an average density of $0.73$ and $1.00$ g$\cdot$cm$^{-3}$ in the decane and water subsystems, respectively, consistent with previous studies\cite{rainer2012,Poli2020NatCommun}. The DPMD simulation was carried out under periodic boundary conditions and in the NVT canonical ensemble for 5 ns. The first ns was employed for equilibration, the last 4 ns were used for analyses. The temperature was kept fixed at $300$~K via the canonical-sampling-velocity-rescaling (CSVR) thermostat~\cite{CSVR} with a damping factor of 0.04 ps. 

1000 randomly chosen configurations were extracted from the DPMD trajectory and used as input for \emph{ab initio} calculations at the r2SCAN \cite{r2SCANerratum} Density Functional Theory (DFT) level. In addition to a series of DFT benchmarks performed on several dimer species and up to the coupled-cluster singles and doubles (CCSD) theory \cite{Cizek69,Purvis82,Scuseria88,Scuseria89}, preliminary tests on 100 extended decane-water configurations extracted from the DPMD simulations were carried out at the BLYP+D3 \cite{BLYP1,BLYP2,Grimme1,Grimme2}, SCAN \cite{SCAN}, and r2SCAN \cite{r2SCANerratum} levels to compare the current results with previous GGA-based evaluations\cite{Poli2020NatCommun} and to understand the impact of the electron delocalization in charge transfer (CT) quantification, as reported in the Supporting Information (SI). Calculations on dimers were performed via the Orca software (v. 6.1) \cite{Orca} in the gas-phase, while all periodic electronic structure calculations of the extended oil-water interface were carried out by employing the CP2K software (v. 2024.3) \cite{CP2K}. Wavefunctions of the atomic species have been expanded in TZVP basis sets with Goedecker-Teter-Hutter (GTH) pseudopotentials using the GPW method~\cite{GTH3}. In addition, a large plane-wave cutoff of 1200 Ry has been imposed, as typically required for r2SCAN calculations.

\subsection{Charge Transfer Analysis}

A simple approach to quantify charge transfer (CT) was proposed by Belpassi and co-workers,\cite{belpassi0,belpassi} based on directly comparing the electron density of a complex with that of its isolated fragments. This method accurately captures CT in weakly interacting systems, such as noble gases interacting with hydrating water, in agreement with molecular beam scattering measurements probing interatomic potentials.\cite{belpassi_PCCP2009,belpassi_JACS2010,cappelletti_AccChemRes2012} As shown in the SI, our approach yields excellent agreement with the Belpassi scheme for CT in the water dimer.\cite{belpassi}

Using a similar approach, we quantify interfacial CT from first-principles electron densities. Specifically, we compute the electron density of the full interacting system and, in the same nuclear geometry and simulation box, the electron densities of the two isolated subsystems (i.e., subsystem~1 and subsystem~2). The difference
\(\Delta\rho(\mathbf r)=\rho_{\mathrm{full}}(\mathbf r)-\rho_{\mathrm{sub1}}(\mathbf r)-\rho_{\mathrm{sub2}}(\mathbf r)\)
defines the interaction-induced density rearrangement, which highlights where electrons accumulate (positive \(\Delta\rho\)) and where they deplete (negative \(\Delta\rho\)). In the case of the oil-water interface slab simulations, subsystem 1 and 2 correspond to the oil and water phase, respectively. This procedure thus directly probes the collective response of the electron density from both subsystems. By construction, the total number of electrons is conserved and the integral of \(\Delta\rho\) over the whole space vanishes.

To attribute the density rearrangement to individual atoms, we partition the real space into atom-centered cells using a van der Waals (vdW)-weighted power diagram, which expands the spatial domain of larger atoms relative to smaller ones in a physically meaningful way (see the SI for details and tests). This is akin to the Voronoi deformation density (VDD) charges used for atoms and molecules in quantum chemistry\cite{matthias2004} which do not directly depend on the use of basis functions. Integrating \(\Delta\rho\) over each atomic cell yields an atom-resolved net electron change \(q_i\), which we report in units of electrons. Throughout, we adopt the convention by which a \emph{positive} CT leads to a \emph{positive} variation of \(q_i\), meaning electron \emph{accumulation} on atom \(i\) (i.e., a more \emph{negative} physical charge \(-e\,q_i\)); conversely, a \emph{negative} CT producing a \emph{negative} \(q_i\) indicates electron \emph{depletion} on atom \(i\). Global charge conservation is satisfied to within a small numerical residual of \(\sim 5\times 10^{-5}\,e^{-}\), which we regard as the numerical error on the CT values.

To map where depleted electrons plausibly flow (i.e., the CT direction), we cast the problem as a simple optimal-transport (minimum-cost assignment) between donor atoms (losing electron density) and acceptor atoms (gaining electron density), in the spirit of the Monge-Kantorovich optimal transport formulation \cite{Villani2009,Cuturi2013}. We introduce a non-negative transport matrix \(T\), whose entries \(T_{ij}\) represent the amount of electron flow from electron donor \(i\in\mathcal D\) to electron acceptor \(j\in\mathcal A\), and determine \(T\) by minimizing the total pairing cost,
\[
\min_{T_{ij}\ge 0}\;\; \sum_{i\in\mathcal D}\sum_{j\in\mathcal A} c_{ij}\,T_{ij},
\qquad
c_{ij}=\frac{d_{ij}}{R_i^{\mathrm{vdW}}+R_j^{\mathrm{vdW}}},
\]
where \(d_{ij}\) is the interatomic distance and \(R_i^{\mathrm{vdW}},\,R_j^{\mathrm{vdW}}\) are the corresponding van der Waals radii, thereby accounting for the different size of the nuclei. The solution is an optimal \(T\)-matrix, whose entries \(T_{ij}\) are the number of electrons reassigned from donor \(i\) to acceptor \(j\). As shown in the SI, such a vdW-weighted Voronoi partitioning applied to the density difference $\Delta \rho(\mathbf r)$ enables a physically consistent assignment of electrons to atoms (Table S1) and quantification of interaction-induced charge rearrangements (Table S2). These contributions are summarized in a compact $2\times2$ electronic $T$-matrix notation, where diagonal elements describe intramolecular self-polarization and off-diagonal elements represent genuine intermolecular CT.

\section{Results and Discussion}

\subsection{Charge Transfer in Dimers}

\begin{figure}
    \centering
    \includegraphics[width=\linewidth]{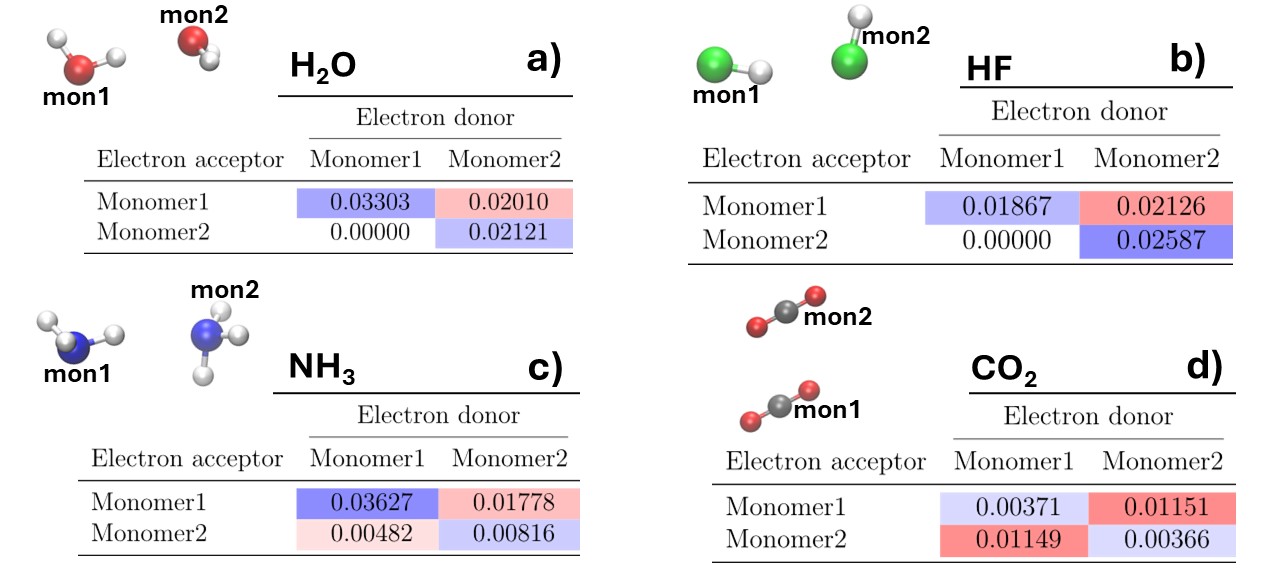}
    \caption{\scriptsize Electronic $T$-matrix of the water (a), hydrogen fluoride (b), ammonia (c), and carbon dioxide (d) dimer. Molecular geometries are optimized at the M06-2X/def2-TZVPD level whereas the respective electron densities are evaluated at the r2SCAN/def2-TZVPD theory. Diagonal elements (blue) correspond to self-polarization while off-diagonal entries (red) correspond to the inter-molecular charge transfer. Net charge transfer is simply obtained by taking the absolute difference between off-diagonal elements. Color intensity of the cells follows the magnitude of the elements within each $T$-matrix while zero values are not colored.}
    \label{figure1}
\end{figure}

As a first validation of our framework, we apply the present methodology to a set of prototypical molecular dimers exhibiting different interaction strengths and bonding motifs. These simple systems provide a controlled benchmark to assess whether our method can discriminate between distinct interaction regimes -- from strong hydrogen bonds to weak van der Waals contacts -- while consistently quantifying both intermolecular charge transfer (CT) and intramolecular polarization. Furthermore, this approach allows us to go beyond the need to identify the isodensity point to partition the electron density,\cite{belpassi} a critical aspect very recently discussed in Ref.~\cite{MHG_JCTC26}. By contrast, the vdW-weighted Voronoi partitioning combined with the optimal-transport formulation introduced here provides an unambiguous description of the electronic rearrangements induced by intermolecular interactions, as illustrated by the $2\times2$ $T$-matrices reported in Fig.~\ref{figure1}.

Strong hydrogen bonding in the water (Fig.~\ref{figure1}a) and hydrogen fluoride (Fig.~\ref{figure1}b) dimers produces a characteristic $T$-matrix pattern where one off-diagonal element reaches $\sim0.020~e^{-}$ ($20~me^{-}$) while the other is essentially zero. As discussed in the SI, this asymmetry reflects partial electron transfer from the hydrogen-bond acceptor (monomer2) to the donor (monomer1)\cite{glendening2005,khaliulin2009}. The diagonal self-polarization terms (blue) show opposite trends in the two systems: in the water dimer the donor is more strongly re-polarized, whereas in HF the acceptor experiences a larger polarization. This difference arises from the interplay between molecular polarizability, bond polarity, and geometry: the nearly orthogonal hydrogen-bond arrangement in the planar HF dimer enhances polarization of the acceptor lone-pair region, while the more cooperative orbital alignment in the water dimer favors stronger re-polarization of the donor unit.

The ammonia dimer is more subtle because its potential energy surface is rather flat, leading to highly floppy structures around the stationary points~\cite{Ammonia_PES}. We therefore analyze the linear staggered structure with C$_s$ symmetry. The hydrogen-bond signature clearly appears in the corresponding $T$-matrix (Fig.~\ref{figure1}c), although the net CT (CT$_{net}=|T_{mon2\rightarrow mon1}-T_{mon1\rightarrow mon2}|\approx0.013~e^{-}$) is smaller than in the H$_2$O and HF dimers, indicating a less efficient hydrogen bond for CT. The NH$_3$ dimer also exhibits asymmetric self-polarization, with the donor fragment undergoing stronger re-polarization (T$_{mon1\rightarrow mon1}$) than the acceptor (T$_{mon2\rightarrow mon2}$). A markedly different pattern arises for the CO$_2$ dimer (Fig.~\ref{figure1}d). Owing to its weak van der Waals interaction and centrosymmetric parallel-slipped C$_{2h}$ structure, the system shows very small and nearly symmetric re-polarization effects. Despite finite CT contributions on each monomer, the net CT essentially vanishes (CT$_{net}\sim2\times10^{-5}~e^{-}$).


\begin{figure}
    \centering
    \includegraphics[width=0.95\linewidth]{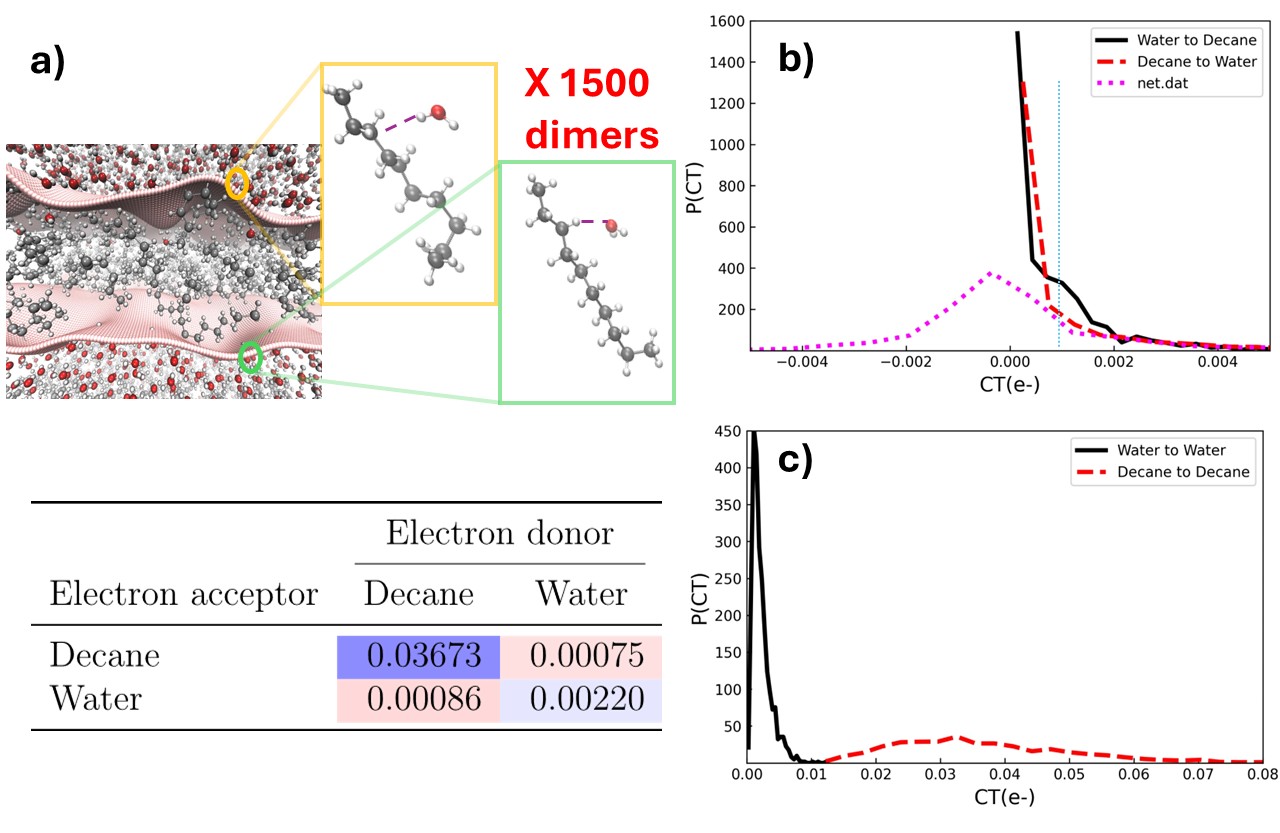}
    \caption{\scriptsize (a) Snapshot taken from the Deep Potential Molecular Dynamics simulation of the extended decane-water interface and the extraction procedure of decane-water dimers along with the respective $T$-matrix averaged over 1500 configurations. (b) Distributions of the charge transfer (CT, in \(e^{-}\)) by direction. Black solid line: water \(\rightarrow\) decane CT; red dashed line: decane\(\rightarrow\)water CT; magenta dotted line: net CT. (c) Distributions of the intra-monomer (self-polarization) CT. Black solid line: water\(\rightarrow\)water; red dashed line: {decane\(\rightarrow\)decane}.  }
    \label{figure}
\end{figure}

Our goal in this work is to examine the CT between decane and water specifically comparing how this effect changes moving from isolated clusters to a realistic extended interface. We first begin by exploring the CT we extract for the decane-water dimers in vacuum. Following a similar strategy to that adopted in Ref.\cite{Zhao2025Angew}, we have extracted 1500 local configurations of decane-water dimers from our Deep Potential Molecular Dynamics (DPMD) trajectory of the extended decane-water interface, as visualized in Fig. \ref{figure}a. We focused on two types of typical configurations found at the oil-water interface as depicted in the yellow and green images of Fig.~\ref{figure}a, corresponding to situations where the water hydrogens are either oriented toward or away from the decane molecule respectively. 

Accumulating all the per-atom charge fluctuations from the atoms involved in this dimer configuration, we observed that the self-polarization associated with decane is markedly larger than the water one, as quantified by the diagonal terms of the $T$-matrix reported in Fig.~\ref{figure}a and by the probability distributions of Fig.~\ref{figure}c. More importantly, an almost symmetric CT between decane and water emerges from this analysis. In fact, the absolute difference of the off-diagonal elements of the $T$-matrix produces an insignificant CT of $0.00011~e^{-}$ ($0.11~me^{-}$) imbalance toward water, as quantified in Fig.~\ref{figure}a and visualized in Fig.~\ref{figure}b by the net CT curve (dotted magenta). This almost-zero net CT between hydrocarbon and water dimers is consistent with recent ALMO-EDA~\cite{ALMO-EDA} simulations by the Head-Gordon group~\cite{Zhao2025Angew} on hexane-water systems. These findings imply that the CT associated with decane-water clusters are essentially insignificant. Next we move to investigating the CT observed at extended oil-water interfaces.

\subsection{Charge Transfer at the Extended Decane-Water Interface}

\begin{figure}
    \centering
    \includegraphics[width=\linewidth]{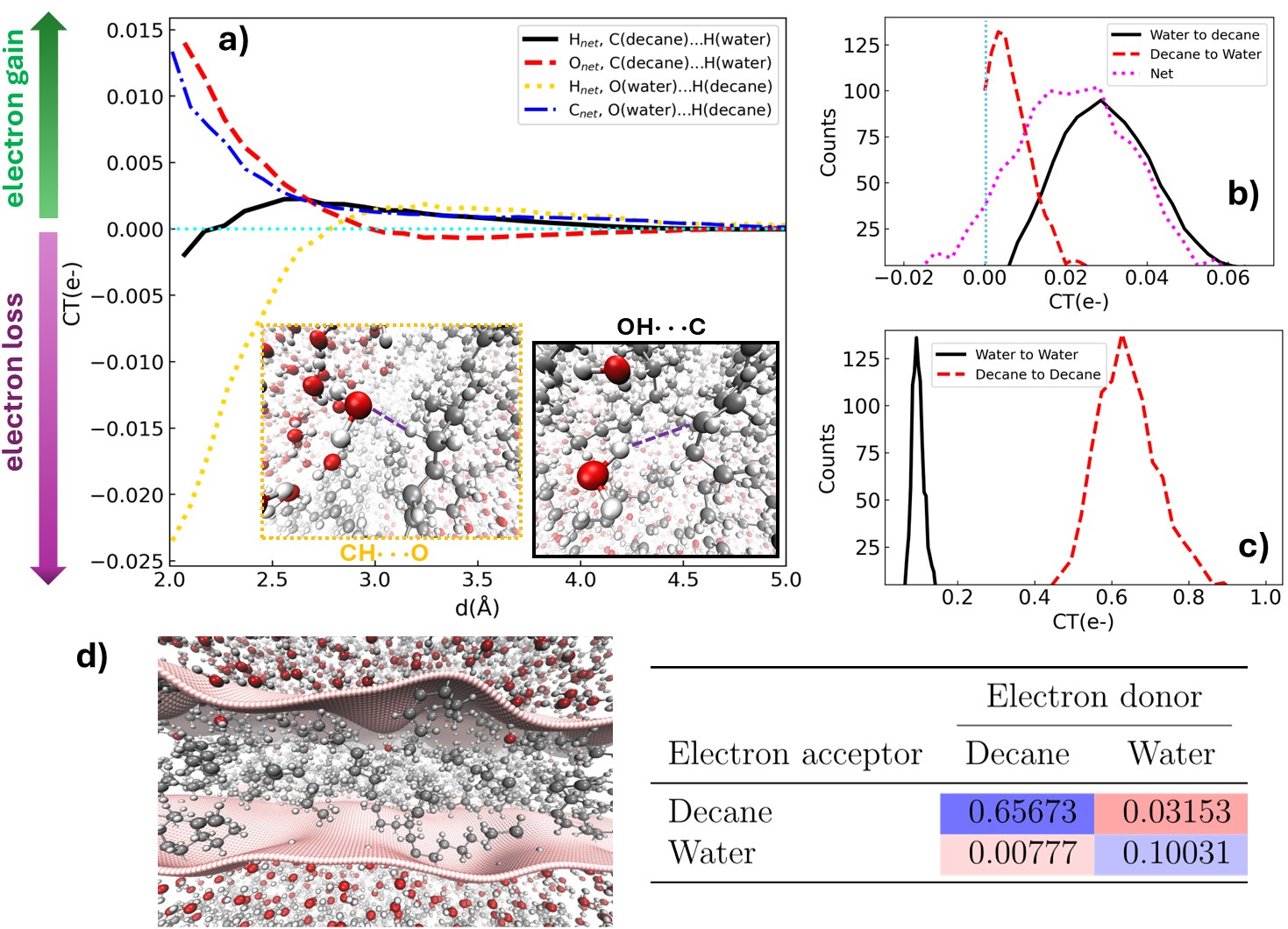}
    \caption{\scriptsize (a) Conditional mean of the net atomic charge, \(\langle q^{\mathrm{net}}\rangle\) (in \(e^{-}\)), plotted as a function of the cross-phase approach distances. Curves: black solid, water H net charge; red dashed, water O net charge; orange dotted, decane H net charge; blue dash-dot, decane C net charge. The thin cyan dotted horizontal line denotes \(q^{\mathrm{net}}=0\). Values are shown as bin-averaged means of \(\langle q^{\mathrm{net}}\rangle\) conditioned on the distance of the interaction. Yellow dotted inset displays the forward interaction (decane to water). Black continuous inset shows the backward interaction (water to decane). (b) Distributions of interfacial charge transfer (CT, in \(e^{-}\)) by direction. Black solid line: water \(\rightarrow\) decane CT ; red dashed line: decane\(\rightarrow\)water CT; magenta dotted line: net CT. (c) Distributions of intra-phase (self-polarization) CT. Black solid line: water\(\rightarrow\)water; red dashed line: decane\(\rightarrow\)decane. (d) Snapshot from a Deep Potential Molecular Dynamics trajectory of the decane-water slab, showing the Willard-Chandler instantaneous interface between the two phases (pale pink dots). Carbon, oxygen, and hydrogen atoms are depicted as silver, red, and white spheres, respectively. On the right, the respective average coarse-grained $2\times2$ $T$-matrix with the electron density re-sampled @r2SCAN/TZVP level. Entries are in $e^{-}$.}
    \label{figure2}
\end{figure}

We begin by quantifying the electronic response that develops upon formation of the decane-water extended interface. Using the electron density decomposition and CT evaluation protocol introduced above, we (i) resolve the atom-specific net charge response as a function of the cross-phase distances, (ii) determine the net CT direction and the amount of the interfacial CT between the two phases, and (iii) isolate the intra-phase self-polarization occurring independently within water and within the hydrocarbon matrix. Fig.~\ref{figure2} summarizes these three complementary results.

Fig.~\ref{figure2}a reports the conditional mean of the net atomic charge lying on the atomic species as a function of the cross-phase approach distance $d$, providing an atom-resolved picture of the different interfacial CT terms. In particular, the four CT curves for the water hydrogens (black solid), the water oxygens (red dashed), the decane hydrogens (yellow dotted), and the decane carbons (blue dash-dotted) are plotted as a function of two different relative distances: the C$_{decane}$-H$_{water}$ (black solid and red dashed lines) and the O$_{water}$-H$_{decane}$ (yellow dotted and blue dash-dotted lines). As expected, all curves converge toward zero at larger, bulk-like separations, confirming that the charge response induced upon interface formation is relatively short-ranged along directions approximately normal to the interface itself. 

Within the water phase along the C$_{decane}\cdots$H$_{water}$ distance ranging between $\sim2-3$~{\AA}, oxygen atoms (red dashed curve) at contact exhibit a partial electron accumulation due to the hydrogen bonds they donate to decane (O--H$\cdots$C), while at farther distances the observed electron density depletion mirrors their involvement as hydrogen-bond acceptors in C--H$\cdots$O interactions. Water hydrogens display the opposite response (solid black curve), exhibiting charge depletion at the topmost interfacial water layer ($\sim2.0-2.2$~{\AA}), where some of the hydrogen atoms of the water subsystem are capable of donating efficient forward hydrogen bonds to the decane moiety. However, the remainder water's hydrogen atoms accumulate some extra electron charge within a relatively large range of distances of interactions established with the oil phase ($\sim2.3-4.1$~{\AA}). The sign of this CT arises from a mixture of cooperative effects including the response from waters losing electron density in C--H$\cdots$O interactions as well as hydrogens in dangling (free O-H's) at the interface.


At close contact with water, carbon atoms on the hydrocarbon side acquire a negative mean charge, indicating electron accumulation on interfacial carbons. Although the short-range CT behavior resembles that of water oxygen atoms, decane carbons remain more negatively charged than their bulk counterparts over the entire interaction range up to $\sim5.0$~{\AA}, as indicated by the positive blue dash-dotted curve in Fig.~\ref{figure2}a. Conversely, the hydrocarbon's hydrogen atoms (yellow dotted line) become electron-poor when directly engaged in cross-phase contacts up to a O$_{water}\cdots$H$_{decane}$ distance of $\sim2.8$~{\AA}, where the respective CT curve crosses zero and changes sign. This indicates that whereas the decane's hydrogen atoms donating hydrogen bonds to water get positively charged, their counterpart lying on the opposite side of the hydrocarbon backbone tends to acquire a partially negative charge, similarly to their covalently bound carbon atoms. Notably, the spatial extent of this partially charged region is larger in the oil phase than in the water, as witnessed by the non-zero values of the H$_{net}$ and C$_{net}$ curves up to $\sim5.0$~{\AA} of the O$_{water}\cdots$H$_{decane}$ distance, while the H$_{net}$ and O$_{net}$ CT curves referred to the aqueous subsystem approach the bulk-like zero net CT already at about $4.0$~{\AA} of the C$_{decane}\cdots$H$_{water}$ distance. 

The directionality of the interfacial charge flow is illustrated in Fig.~\ref{figure2}b, which reports the distributions of the CT between each subsystem. By construction, the charge donated by each subsystem is a positively defined quantity, since no atom can donate more electrons than it holds. The amount of CT transferred from decane to water (red dashed curve) attains a maximum value of $\sim0.020~e^{-}$ (20~$me^{-}$). In contrast, the forward CT distribution from water to decane (black solid curve) is shifted toward larger values than the backward CT from decane to water, resulting in a positive net CT with an average of $0.024~e^{-}$ (24~$me^{-}$) and fluctuations reaching up to $0.060~e^{-}$ (60~$me^{-}$). These values are quantified in the average $2\times2$ $T$-matrix reported in Fig. \ref{figure2}d, where a net average CT of $0.02376~e^{-}$ from water to decane emerges. This imbalance indicates \emph{an overall accumulation of electronic charge on the hydrocarbon side of the interface} induced by the interaction with water, a feature that is not captured when the analysis is conducted on water-decane dimer clusters (Fig. \ref{figure}).

Earlier we saw that for the decane-water dimer systems, a significant part of the electronic response involves the self-polarization contribution of the decane and water molecules. Fig.~\ref{figure2}c isolates the intra-phase self-polarization induced upon interface formation. The hydrocarbon self-polarization distribution is centered at substantially larger values ($\sim0.6$--$0.7~e^{-}$) than water ($\sim0.1$--$0.15~e^{-}$), indicating that the decane phase accommodates the interfacial perturbation through a much stronger internal charge rearrangement, which is approximately one order of magnitude larger than that exhibited by water. This is of course rooted in the fact that the hydrocarbon is a larger molecule and therefore with more electronic degrees of freedom available to be perturbed by the creation of the interface.


Previous studies of CT mechanisms at air-water and oil-water interfaces suggest that interfacial charge accumulation stems from asymmetries in the populations of coordination defects with differing numbers of hydrogen-bond donors and acceptors\cite{Vacha2011JACS,Poli2020NatCommun}. For the hexane-water interface, Zhao et al.~\cite{Zhao2025Angew} showed that backward and forward CT balance each other leading to essentially no net charge displacement at the interface. To understand the origins of the CT at the decane-water interface, we examined the different possible orientations of water molecules relative to the instantaneous Willard-Chandler (WC) surface.\cite{Willard2010InstantaneousInterface} Specifically, atoms are classified as interfacial when their distance from the WC surface satisfies $|d_{\mathrm{WC}}|\le d_{\mathrm{phase}}$, with $d_{\mathrm{C}}=2.8$ {\AA} for decane carbons and $d_{\mathrm{W}}=3.0$ {\AA} for water oxygens. This labeling allows us to distinguish two interaction motifs: the forward C-H$\cdots$O and the backward O-H$\cdots$C contacts, characterized by the corresponding approach distances and angles. Within our CT convention, the hydrogen-bond donor corresponds to the net electron-accepting side, so that the C-H$\cdots$O motif correlates with CT from water to decane, and viceversa.

\begin{figure}[t!]
\centering
\includegraphics[width=\linewidth]{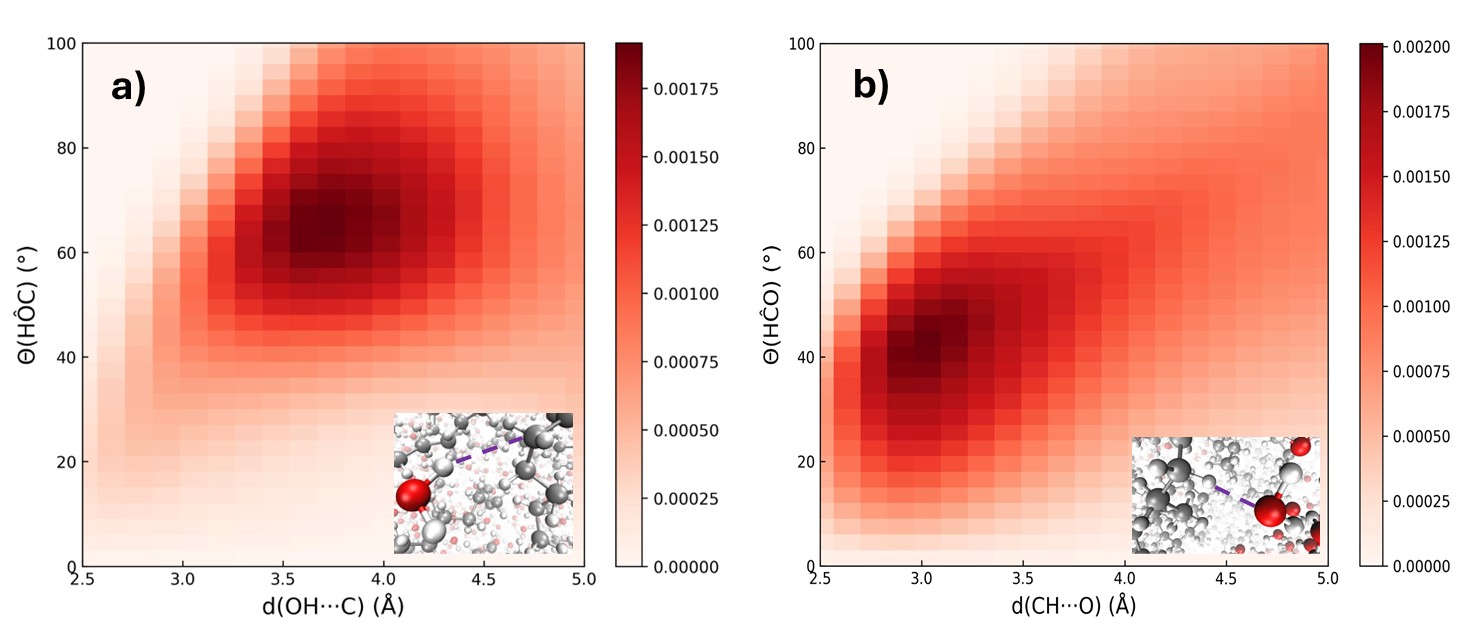}
\caption{\scriptsize (a) Two-dimensional probability density $P\!\left(d_{\mathrm{OH\cdots C}},\,\theta_{\mathrm{HOC}}\right)$, where $d_{\mathrm{OH\cdots C}}\equiv d(\mathrm{OH}\cdots\mathrm{C})$ and $\theta_{\mathrm{HOC}}\equiv\angle(\mathrm{H}_{\mathrm{water}}-\mathrm{O}-\mathrm{C})$ of the backward O-H$\cdots$C interaction motif (water $\rightarrow$ decane). 
(b) Two-dimensional probability density $P\!\left(d_{\mathrm{CH\cdots O}},\,\theta_{\mathrm{HCO}}\right)$, where $d_{\mathrm{CH\cdots O}}\equiv d(\mathrm{CH}\cdots\mathrm{O})$ and $\theta_{\mathrm{HCO}}\equiv\angle(\mathrm{H}_{\mathrm{decane}}-\mathrm{C}-\mathrm{O})$ of the forward C-H$\cdots$O interaction motif (decane $\rightarrow$ water). Only atoms tagged as interfacial via the instantaneous Willard-Chandler surface are sampled, using thresholds $s_{\mathrm{C}}=2.8~\text{\AA}$ (carbons) and $s_{\mathrm{W}}=3.0~\text{\AA}$ (oxygens). Insets show representative configurations used to define the geometric descriptors.}
\label{Figure3}
\end{figure}

Fig.~\ref{Figure3} indeed shows a clear asymmetry between the two cross-phase hydrogen-bond-like interactions. In Fig.~\ref{Figure3}a, the backward O--H$\cdots$C motif spans a broader region shifted toward longer distances and larger $\theta_{\mathrm{H\hat{O}C}}$ angles, consistent with weaker and less directional interactions. In contrast, the forward C--H$\cdots$O motif (Fig.~\ref{Figure3}b) samples shorter distances and $\theta_{\mathrm{H\hat{C}O}}$ angles, indicating stronger and more directional contacts. Because CT scales as $e^{-r}$ with the interaction distance $r$, this implies that CT from water to decane is substantially more efficient than in the opposite direction, consistent with the analysis of Fig.~\ref{figure2}. This geometrical asymmetry provides a microscopic basis for the electronic imbalance identified above: the shorter and more aligned C--H$\cdots$O contacts constitute a dominant channel for interfacial polarization and for the emergence of a net CT bias at oil-water interfaces.



\subsection{Structural Fingerprints of Charge-Transfer at Oil-Water Interface}

The analysis of the change in electron density underlying the creation of an extended decane-water interface reveals a collective process involving electronic polarization within the decane and water molecules as well as between the two phases. These mechanisms cannot be directly probed experimentally and are therefore typically inferred indirectly from surface sensitive vibrational spectroscopy\cite{StrazdaiteJCP2015,Pullanchery2021Science,YangPRL2020} which arise due to a geometrical response of the nuclear degrees of freedom. Since our CT analysis examines the electronic response at fixed nuclear geometry, we turn to examining structural properties of the two phases from the DPMD simulations which have effectively learned the electronic properties such as the energies and forces. Without explicitly computing relevant IR or Raman spectra, we can only make indirect suggestions on possible vibrational responses one might expect at the oil-water interface.


\begin{figure}[t!]
\centering
\includegraphics[width=\linewidth]{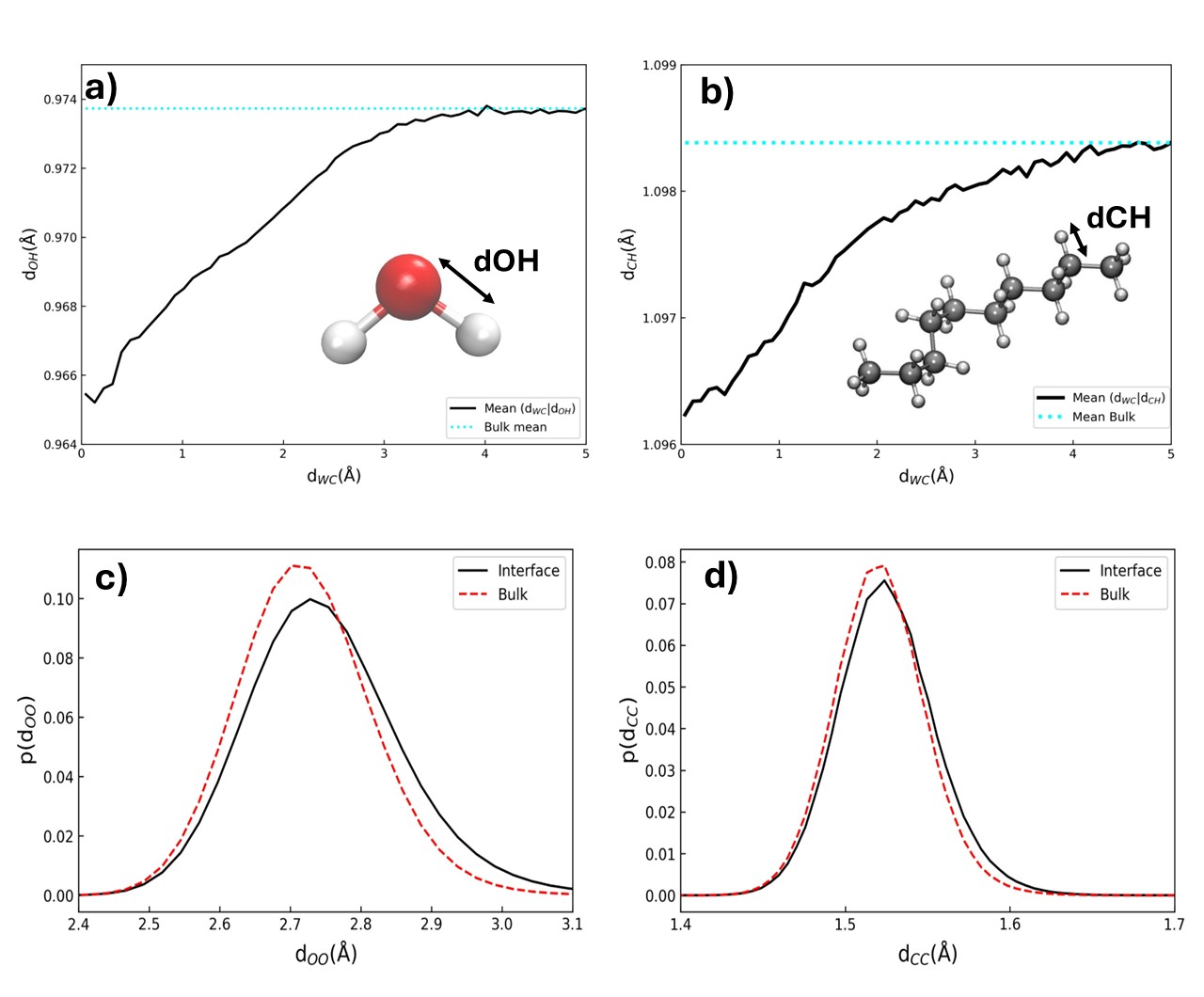}
\caption{\scriptsize (a) Conditional mean of the covalent O--H bond length of water molecules, $\langle d_{\mathrm{OH,cov}}\mid d_{\mathrm{WC}}\rangle$, as a function of the distance to the instantaneous WC surface $d_{\mathrm{WC}}$ (black solid line). The cyan dotted line marks the bulk mean $\langle d_{\mathrm{OH,cov}}\rangle_{\mathrm{bulk}}$.
(b) Conditional mean of the covalent C--H bond length of decane molecules, $\langle d_{\mathrm{CH,cov}}\mid d_{\mathrm{WC}}\rangle$, as a function of $d_{\mathrm{WC}}$ (black solid line), compared to the bulk mean $\langle d_{\mathrm{CH,cov}}\rangle_{\mathrm{bulk}}$ (cyan dotted line).
(c) Probability density of the nearest-neighbor O$\cdots$O distance $d_{\mathrm{OO}}$ in water: interface (black solid) vs bulk (red dashed).
(d) Probability density of the nearest-neighbor C$\cdots$C distance $d_{\mathrm{CC}}$ in decane: interface (black solid) vs bulk (red dashed).}
\label{Figure4}
\end{figure}

Fig.~\ref{Figure4} shows that the interface induces subtle but systematic and measurable modifications within both phases. On the water side (Fig.~\ref{Figure4}a), the conditional mean $\langle d_{\mathrm{OH,cov}}\mid d_{\mathrm{WC}}\rangle$ -- which represents the mean O--H covalent bond distance of the water molecules belonging to the interfacial region -- is reduced at small distances from the Willard-Chandler (WC) interface, $d_{\mathrm{WC}}$, and gradually approaches the bulk mean upon moving away from the interface. This indicates a slight contraction of the O--H covalent bonds in the interfacial region. Concomitantly, the $d_{\mathrm{OO}}$ distribution monitoring the intermolecular distances between water molecules (Fig.~\ref{Figure4}c) is shifted toward larger separations at the interface, consistent with a mild opening of the first coordination shell with respect to the typical bulk-like condition. Taken together, these structural changes would imply that the vibrational frequencies of the O-H oscillators at the interface would be slightly blue-shifted in frequency relative to those in the bulk. This likely originates from the presence of dangling O-H bonds from an enhanced concentration of defects\cite{Poli2020NatCommun}.



Structural changes in the water molecules do not directly probe the complex interactions at the interface. We thus turned next to examining the structural coordinates of the decane molecules. Although C--H bonds are less polarizable than O--H bonds, Fig.~\ref{Figure4}b reveals a small decrease of the conditional mean $\langle d_{\mathrm{CH,cov}}\mid d_{\mathrm{WC}}\rangle$ -- which represents the mean C--H covalent bond distance of the decane molecules belonging to the interfacial region -- close to the WC surface, which relaxes toward the bulk value at larger $d_{\mathrm{WC}}$. Interestingly, this effect is consistent with experimental indications of interface-perturbed C--H bonds and the blue-shift of the respective stretching mode~\cite{Pullanchery2021Science}. At the same time, the nearest-neighbor C$\cdots$C distribution (Fig.~\ref{Figure4}d) is slightly shifted and broadened toward larger distances at the interface, suggesting a modest loosening of the hydrocarbon backbone packing and reflecting the negatively charged nature of the carbon atoms at the interface previously discussed.
Notably, the fact that both $d_{\mathrm{OO}}$ and $d_{\mathrm{CC}}$ respond in the same direction -- i.e., toward slightly larger first-neighbor separations -- suggests that the interface promotes a modest local “opening” of the nearest-neighbor structure in each phase. Within this framework, the interface emerges as a region of enhanced structural and electronic susceptibility, where subtle but measurable net inter-phase CT contributions results in non-local polarization and structural rearrangements. These effects appear to be at the basis of the observed vibrational shifts of specific modes at oil-water interfaces.



\section{Conclusions}

In this work, we have presented a quantitative, many-body, electronic-structure-based characterization of charge rearrangements at an extended oil-water interface by combining Deep Potential Molecular Dynamics, first-principles electronic densities, and a real-space, atom-resolved charge-partitioning and optimal-transport analysis. This framework allows us to disentangle inter-phase charge transfer (CT) from intra-phase self-polarization and to directly compare the electronic response of isolated dimers with that of a realistic, fluctuating interface. 

Our results show that decane-water dimers exhibit nearly symmetric forward and backward CT, leading to an almost vanishing net CT in agreement with recent energy-decomposition analyses. In striking contrast, the extended interface displays a clear electronic asymmetry, yielding a systematic net CT from water to the hydrocarbon phase. We attribute this net CT to the tendency for water molecules to accept weak improper hydrogen-bonds from the C-H groups. The collective effect of the electronic polarization involving both intra-and-inter phase responses, is also reflected in structural changes at the interfaces in particular, the contraction of C-H bonds resulting in specific vibrational spectroscopic signatures.


The surface charge densities that we obtain ($\sigma=0.00624~e^{-}\,\mathrm{nm}^{-2}\sim1.0~\mathrm{mC\,m^{-2}}$) are up to one order of magnitude larger than those inferred from previous simulations with extended hexane-water interfaces where CT was only empirically included between water molecules\cite{Vacha2011JACS} as well as the ALMO/EDA CT analysis reported for hexane-water clusters sampled from interface simulations \cite{Zhao2025Angew}. On the other hand, our values are about a factor of two-three smaller than those previously reported by Poli and Hassanali ($0.015~e^{-}\,\mathrm{nm}^{-2}$)~\cite{Poli2020NatCommun}. This discrepancy very likely originates from the tendency of GGA functionals like BLYP to overdelocalize charge\cite{delocalization_error} (see SI for details).

As eluded to earlier in the introduction, there is currently a lot of experimental and theoretical interest about the possibility of enhancing chemical reactivity at interfaces\cite{RuizLopez2020AqueousInterfaces}. While this still remains a highly contentious topic, one of the popular invoked arguments to rationalize these phenomena are the presence of large electric fields which are thought to lower the barriers for chemical reactions. In fact, very recently, large electric fields at oil-water interfaces (50-90~MV~cm$^{-1}$)~\cite{Shi2025Nature} have also been inferred using a combination of both surface sensitive Raman spectroscopy and molecular simulations. Our net CT estimate instead suggests that substantially weaker fields are present under the conditions explored here. Assuming a simple continuum electrostatics model, we determine that the approximate value of the electric field created by the CT is $\sim 0.014-0.3$~MV~cm$^{-1}$, which is about 3-4 orders of magnitude smaller than previous estimates~\cite{Shi2025Nature}. While our predictions are in line with second harmonic scattering experiments, which yield a field strength upper bound of $\sim 0.1$~MV~cm$^{-1}$, \cite{PullancheryNatCommun2024} it should be stressed that this electric field we infer does not include other effects including dipolar/quadrupolar contributions and thus the agreement maybe fortuitous.

Taken together, our findings demonstrate that the charge state of oil-water interfaces is an emergent, collective property of the extended interface, governed by many-body polarization and hydrogen-bond-network asymmetries rather than by local dimer physics alone. This work provides a unified electronic-structure perspective that reconciles cluster-based and interfacial viewpoints and offers a robust framework for interpreting electrostatics and spectroscopy at complex hydrophobic surfaces in contact with aqueous systems.

\begin{acknowledgement}

G.~C. acknowledges Leonardo Belpassi for insightful discussions on charge transfer. G.~A. and G.~C. acknowledge support by ICSC - Centro Nazionale di Ricerca in High Performance Computing, Big Data and Quantum Computing, funded by European Union - NextGenerationEU - PNRR, Missione 4 Componente 2 Investimento 1.4. G.~C. acknowledges the European Union - NextGeneration EU  from the Italian Ministry of Environment and Energy Security POR H2 AdP MMES/ENEA with involvement of CNR and RSE, PNRR - Mission 2, Component 2, Investment 3.5 ``Ricerca e sviluppo sull’idrogeno'', CUP: B93C22000630006. G.~C. acknowledges the  European Union (NextGeneration EU), through the MUR-PNRR project SAMOTHRACE (ECS00000022). G.~C. is thankful to CINECA for awards under the ISCRA initiative, for the availability of high-performance computing resources and support. A.H. acknowledges funding from the European Research Council (ERC) under the European Union’s Horizon 2020 research and innovation programme (grant agreement No. 101043272 – HyBOP). The views and opinions expressed are those of the authors only and do not necessarily reflect those of the European Union or the European Research Council Executive Agency. Neither the European Union nor the granting authority can be held responsible for them.

\end{acknowledgement}

\begin{suppinfo}


\end{suppinfo}


\bibliography{achemso-demo}

\end{document}